\documentclass[12pt,a4paper]{article}
\usepackage{amsmath}
\usepackage[dvips]{graphicx}
\usepackage{times}
\begin{document}

\begin{titlepage}
\title{Reflection and dips in elastic scattering}
\author{S.M. Troshin,
 N.E. Tyurin
 \\[1ex]
\small  \it Institute for High Energy Physics,\\
\small  \it Protvino, Moscow Region, 142281, Russia}
\normalsize
\date{}
\maketitle

\begin{abstract}
We discuss how the  reflective scattering would affect
elastic scattering in the region of small and moderate values of $-t$,
in particular, we demonstrate that diffractive pattern in the angular distribution
 will be kept  in  a
modified form.
\end{abstract}
\end{titlepage}

Diffraction is a fascinating subject in the light of coming
experiments at the LHC. The term diffraction was introduced to the
hadron and nuclei scattering with use of optical analogy. This
adoption was based on the striking similarities  observed in
hadron and nuclei scattering and light diffraction by  absorbing
obstacles. The absorption in hadron and nuclei scattering is
considered to be a result of opening many inelastic channels at
high energies. Angular distribution  in this absorptive approach
has typical diffraction pattern with prominent forward peak and
secondary maxima and minima and can be associated with wave
properties of particle scattering.

In the recent paper \cite{int} we considered saturation of the
unitarity condition for  scattering matrix
 in hadron collisions at small impact parameters,
when scattering acquires  reflective nature, i.e. $S(s,b)|_{b=0}\to -1$ at $s\to\infty$.
Approach
 to the full
absorption in head-on collisions ---
 the limit $S(s,b)|_{b=0}\to 0$ at $s\to\infty$ --- does not follow
from unitarity itself and is merely  a result of the assumed
saturation of the black disk limit.
On the other hand, the reflective scattering
is a natural interpretation of the unitarity saturation  based on the optical concepts in
high energy hadron scattering.
 Such reflective scattering can be traced to the
continuous  increasing density of
the scatterer with energy, i.e. when density goes beyond some critical value relevant
for the black disk limit saturation,
 the scatterer starts to
acquire  a reflective ability. The concept of reflective
scattering itself is quite general,
 and results from the $S$-matrix  unitarity saturation
related to  the necessity to provide the total cross section
growth at $s\to\infty$. This picture predicts
  that the scattering amplitude at
  the LHC energies is beyond the black disk limit at small impact
  parameters.  The prediction for the total, elastic and
  inelastic cross-sections have been discussed in \cite{int}.

  Reflective scattering should have
  also consequences for the differential cross-section of elastic scattering,
  in the region of small and moderate values of $-t$.
  In principle, it is not evident that
  the presence of the reflective scattering would lead to diffractive angular distributions with
  diffraction peak followed by dips
  and bumps.
It will be shown further on that diffraction pattern will be kept
in a slightly modified form
  in this case.

We start with unitarity condition for the elastic scattering amplitude $F(s,t)$
 which can be written in the form
 \begin{equation}\label{un}
 \mbox{Im}F(s,t)=H_{el} (s,t)+H_{inel} (s,t),
 \end{equation}
 where $H_{el,inel}(s,t)$ are the corresponding elastic  and inelastic overlap function
 introduced by Van Hove \cite{vanh}. Physical meaning of each term in Eq. (\ref{un})
  is evident from the graphical representation in Fig. 1.
\begin{figure}[hbt]
\begin{center}
\includegraphics[scale=0.7]{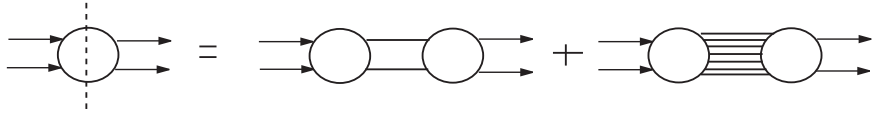}
\end{center}
\caption{\small\it Eq. \ref{un} in the graphical form.}
\end{figure}
The functions $H_{el,inel}(s,t)$ are related to the functions
$h_{el,inel}(s,b)$ and via the Fourier-Bessel transforms, i.e.
\begin{equation}\label{hel}
H_{el,inel} (s,t)=\frac{s}{\pi^2}\int_{0}^{\infty} bdb h_{el,inel}(s,b) J_0(b\sqrt{-t}).
\end{equation}
The elastic and inelastic cross--sections can be obtained as
follows:
\begin{equation}\label{selin}
\sigma_{el,inel}(s)\sim \frac{1}{s} H_{el,inel} (s,t=0).
\end{equation}
As it was already noted, the reflective scattering appears
naturally
 in the $U$--matrix form of unitarization.
In the $U$--matrix approach,
 the $2\to 2$ scattering matrix element in the
impact parameter representation
is the following linear fractional transform:
\begin{equation}
S(s,b)=\frac{1+iU(s,b)}{1-iU(s,b)}. \label{um}
\end{equation}
 $U(s,b)$ is the generalized reaction matrix, which is considered to be an
input dynamical quantity. The relation (\ref{um}) is one-to-one transform and easily
invertible.
Inelastic overlap function $h_{inel}(s,b)$
is connected with $U(s,b)$ by the relation
\begin{equation}\label{hiu}
h_{inel}(s,b)=\frac{\mbox{Im} U(s,b)}{|1-iU(s,b)|^{2}},
\end{equation}
and the only condition to obey unitarity
 is $\mbox{Im} U(s,b)\geq 0$. Elastic overlap function is related to the function
 $U(s,b)$ as follows
\begin{equation}\label{heu}
h_{el}(s,b)=\frac{|U(s,b)|^{2}}{|1-iU(s,b)|^{2}}.
\end{equation}
The form of $U(s,b)$ depends on the particular model assumptions,
but for our qualitative
 purposes it
is sufficient that it increases with energy in a power-like way
 and decreases with impact parameter
like a linear exponent or Gaussian.
To simplify the qualitative picture, we consider also the function $U(s,b)$
as  a pure imaginary.
At sufficiently  high energies ($s>s_0$),
the two separate  regions of
 impact parameter distances can be anticipated, namely the outer region
of peripheral collisions where the scattering has a typical absorptive origin, i.e.
$S(s,b)|_{b>R(s)}>0$ and
 the inner region of central collisions
where the scattering has a combined reflective and absorptive origin, $S(s,b)|_{b< R(s)}<0$.
 The transition to the negative
values of $S$ leads to
appearance of the real part of the phase shift, i.e. $\delta_R(s,b)|_{b< R(s)}=
\pi/2$ \cite{int}. It should be noted here that the quasi-eikonal
form of unitarization allows smooth transition to the reflective scattering mode also.
Unitarity condition   can be obeyed then (beyond the inelastic threshold only)
for a special class of the eikonal functions \cite{sel}.

In what follows we discuss the impact parameter
profiles of elastic and inelastic overlap functions for the scattering picture with reflection
 and for the widely accepted absorptive scattering approach
 in order to make  a conclusion on the absence
 or presence of the dips and bumps in the differential cross-section of elastic scattering,
 which are associated with the zeroes of $\mbox{Im} F(s,t)$ (cf. \cite{giff}
 and references therein).

 We start with the well known absorptive picture of high energy scattering corresponding to the black
 disk limit  at $s\to\infty$. This limit   has already been reached in the head-on
  proton--antiproton
  collisions at Tevatron, i.e. $h_{el}(s,b=0)\simeq 0.25$ \cite{gir}. Thus, one can expect that
  in the framework of absorptive picture, black disk limit will be reached also at $b\neq 0$ at higher
  energies and the impact parameter profiles of $h_{el}(s,b)$ and $h_{inel}(s,b)$
  will be similar  and have a form close to  step function (Fig. 2).
\begin{figure}[hbt]
\begin{center}
\includegraphics[scale=0.6]{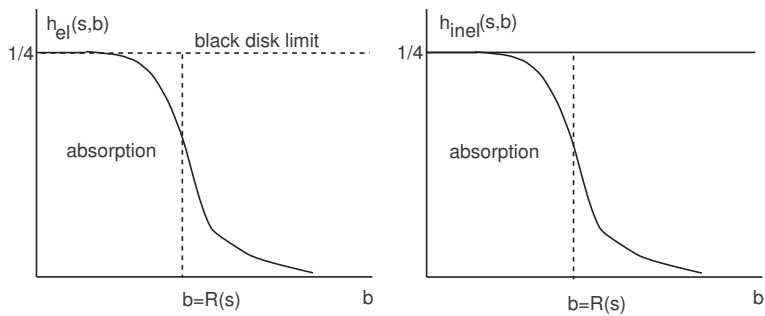}
\end{center}
\caption{\small\it Typical picture of impact parameter profiles of the
elastic and inelastic overlap function in the absorptive approach at asymptotical
 energies.}
\end{figure}
The elastic and inelastic overlap functions $H_{el}(s,t)$ and $H_{inel}(s,t)$
will have similar dependence on $-t$ which can be approximately described as
\begin{equation} \label{sel}
H_{el, inel}(s,t) \sim \frac {RJ_1(R\sqrt{-t})}{\sqrt{-t}}.
\end{equation}
Zeroes and maxima of the functions $H_{el}(s,t)$ and $H_{inel}(s,t)$ are located at the same
values of $-t$, they  will not compensate
each other.  As a result differential cross--section of elastic scattering
 will have well known form wityh dips and bumps
\begin{equation} \label{del}
\frac{d\sigma}{dt} \sim \frac {R^2J_1^2(R\sqrt{-t})}{{-t}}
\end{equation}
when the absorptive scattering picture is realized.
In this case
\begin{equation} \label{tsel}
\sigma_{el,inel}(s)\sim R^2(s).
\end{equation}
 Of course, presence of real part of the scattering amplitude and/or contributions from
 helicity flip amplitudes \cite{dbfl} can modify
the diffraction picture and bring the uncertainty into the above conclusion, but we can suppose
 that those effects are not significant at very high energies.

Let us consider now energy evolution of the elastic and inelastic overlap functions
in the scattering picture which includes reflective scattering.
 With
conventional parameterizations of the $U$--matrix
 the inelastic overlap function increases with energies
at modest values of $s$. It reaches its maximum value $h_{inel}(s,b=0)=1/4$ at some
energy $s=s_0$ and beyond this energy the  reflective
scattering mode appears at small values of $b$. The region of energies and
impact parameters corresponding
to the reflective scattering mode is determined by the conditions
$h_{el}(s,b)> 1/4$ and $h_{inel}(s,b)< 1/4$. The unitarity limit and black disk
limit are the same for the inelastic overlap function, but these limits are different
for the elastic overlap function.
The quantitative analysis of the experimental data
 \cite{pras} gives the threshold value: $\sqrt{s_0}\simeq 2$ TeV.
The function $h_{inel}(s,b)$ becomes peripheral when energy increases
in the region $s>s_0$.
At such energies the inelastic overlap function reaches its maximum
 at $b=R(s)$ where  $R(s)\sim \ln s$.
So, in the energy region, which lies beyond the transition threshold,
 there are two regions in impact
 parameter space: the central region
of reflective scattering combined with absorptive scattering
at $b< R(s)$ and the peripheral region
of pure absorptive scattering at $b> R(s)$. Typical pattern of the impact
parameter profiles for elastic and inelastic overlap functions in the scattering
picture with reflection at the LHC energies is depicted on Fig. 3.
\begin{figure}[hbt]
\begin{center}
\includegraphics[scale=0.6]{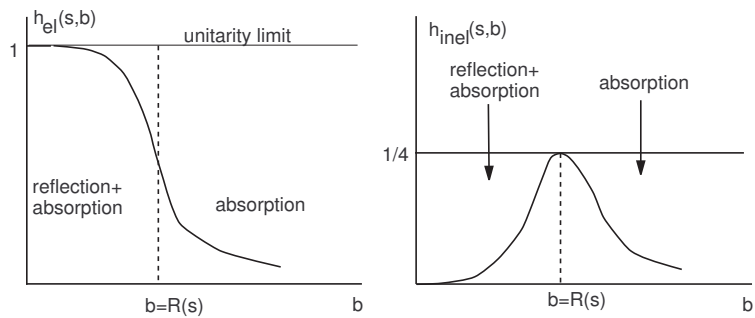}
\end{center}
\caption{\small\it Typical qualitative picture of impact parameter profiles of the
elastic and inelastic overlap function in the reflective approach at the
 asymptotically high energies.}
\end{figure}
Elastic and inelastic overlap functions $H_{el}(s,t)$ and $H_{inel}(s,t)$
will also have different dependencies on $-t$. They can be approximately described as
following
\begin{equation} \label{sela}
H_{el}(s,t) \sim \frac {RJ_1(R\sqrt{-t})}{\sqrt{-t}},
\end{equation}
but
\begin{equation} \label{selaa}
H_{inel}(s,t) \sim RJ_0(R\sqrt{-t})
\end{equation}
and
\begin{equation} \label{tasel}
\sigma_{el}(s)\sim R^2(s),\;\;\; \sigma_{inel}(s)\sim R(s).
\end{equation}
In another words, $H_{el}(s,t)$ dominates over $H_{inel}(s,t)$ at $-t=0$, but it is not the
case for the scattering in the non-forward directions. In this region these
two functions have similar energy dependencies proportional to $R^{1/2}(s)$ at
rather large fixed values of $-t$.
The mean  impact parameter values for elastic and inelastic interactions have also similar
energy dependencies
\begin{equation} \label{taseln}
{\langle b^2\rangle}_{el}(s)\sim R^2(s),\;\;\; {\langle
b^2\rangle}_{inel}(s)\sim R^2(s),
\end{equation}
but the   value of impact parameter averaged over all interactions
\[
{\langle b^2\rangle}_{tot}(s)=\frac{\sigma_{el}(s)}{\sigma_{tot}(s)}{\langle b^2\rangle}_{el}(s)+
\frac{\sigma_{inel}(s)}{\sigma_{tot}(s)}{\langle b^2\rangle}_{inel}(s)
\]
acquires the main contribution  from elastic scattering according to Eq. {\ref{tasel}}.
Therefore, the inelastic intermediate states will give subleading  contribution
to the slope of diffraction cone $B(s)$,
\[
B(s)\equiv \frac{d}{dt}\ln (\frac {d\sigma}{dt})|_{t=0},
\]
at asymptotical energies.
Indeed, since
$B(s) \sim {\langle b^2\rangle}_{tot}(s)$,
it can be written in the form
\[
B(s)=B_{el}(s)+B_{inel}(s),
\]
where
$B_{el}(s)\sim R^2(s)$,
 while $B_{inel}(s)
 \sim R(s)$. It should be noted that
  both terms $B_{el}(s)$ and $B_{inel}(s)$ are
 proportional to $R^2(s)$ in case of the absorptive scattering.

Thus, in the reflective scattering behavior of the function
$H_{inel}(s,t)$ is determined by a peripheral impact parameter
profile and its $-t$ dependence is different. Meanwhile,
 the elastic overlap function $H_{el}(s,t)$
  has similarities with that function in
the case of absorptive approach dependence.  As a result, zeroes
and maxima of the functions $H_{el}(s,t)$ and $H_{inel}(s,t)$
 will be located at different
values of $-t$ and zeroes and maxima of $\mbox{Im}F(s,t)$ will  also  be located
at different position in the cases of absorptive and the reflective scattering.
In the case of reflective scattering, dips and maxima will be  located in the region of
lower values of $-t$.
 We would like to note that the presence of reflective scattering enhances
 the large $-t$ region by factor
  $\sqrt{-t}$ compared to absorptive scattering.
 Despite that these two mechanisms lead at the asymptotics to
 the significant differences in the total, elastic and
 inelastic cross-section dependencies, their predictions for the differential cross-section
 of elastic
  scattering are  not so much different at small and moderate values of $-t$.

We do not consider here the specific model predictions for the differential cross-sections;
our aim was to demonstrate that diffraction picture with dips and bumps in the differential
cross-section will be kept in the case when the reflective scattering dominates.

\section*{Acknowledgements}
We are grateful to V.A. Petrov for the helpful discussions.

\end{document}